\documentclass[11pt]{article}
\usepackage{moriond,epsfig}

\def\be{\begin{equation}}
\def\ee{\end{equation}}
\def\bea{\begin{eqnarray}}
\def\eea{\end{eqnarray}}

\begin{document}
\vspace*{4cm}
\title{Cosmic multi-muon bundles measured at DELPHI \footnote{talk presented in Young Scientists Session of XXXVII th Rencontres de
   Moriond - Electroweak
    Interactions and Unified Theories for  DELPHI
    collaboration}}
\author{ Petr Travnicek, Jan Ridky}
\address{Institute of Physics, Academy of Sciences of the Czech
  Republic}
\maketitle
\abstracts{
 The DELPHI detector at LEP, located $100$ $m$ underground,
 has been used to detect the multi-muon bundles by cathode readout of 
 its hadron calorimeter and its tracking detectors (TPC, muon
 chambers). The experimental apparatus allows us to study muon bundles
 originating from primary cosmic particles with energies in the 
 interval $10^{14}$ - $10^{17}\ eV$. 
 The cosmic events registered during the years 1999 and 2000
 correspond roughly to $1.6\ 10^6 \ s$ of effective run time. 
 The aim of the work is to compare the measured muon multiplicity
 distributions and predictions of high energy interaction
 models for different types of primary particles and also to
 determine the absolute flux of events in certain muon multiplicity
 range. The presentation describes the current status of the analysis.
}
\section{Introduction}
 Cosmic multi-muon bundles are products of atmospheric showers,
 initiated by cosmic rays
 interacting in the upper atmosphere.
% Large detection arrays on the Earth Surface (KASCADE,
% EASTOP, \dots) and underground experiments (MACRO, FREJUS, \dots)
% are used to explore the cosmic ray energy 
% spectrum and to understand
% the  mass composition of primary cosmic nuclei , especially in the
% energy range $10^{15}$~-~$10^{18}$~$eV$.
 The Monte Carlo models ( NEXUS,
 QGSJET, SIBYLL, \dots ) which simulate the nucleus-nucleus interaction
 are used by cosmic ray experiments 
 to correlate measured quantities with primary particle 
 energy and type. These
 models are tuned to data  
 from accelerator detectors  at much lower energies.
 Even if the large progress has been obtained in this field 
 during last years  (eg. KASKADE results ~\cite{kas})
 it is clear that the models should be tuned to as many measurements
 as possible.
 The data presented here were recorded by the DELPHI experiment (CERN)
 at an intermediate underground depth ($100$ $m$) corresponding to
 energy cutoff for vertical muons $\sim 50$ $GeV$.  The 
 study  of multi-muon bundles in this underground depth is
 motivated by observed excess of high multiplicity events in the COSMO-ALEPH
 experiment~\cite{eg}. The independent cross-check of interaction
 models is the main aim of this work.
\section{Data and Simulation}
 The data were taken during years $1999$ and $2000$ in the parasitic
 mode of $e^+e^-$ data taking as well as in dedicated cosmic runs.
 The muon tracks
 have  been reconstructed from barrel part of 
 DELPHI Hadron Calorimeter (HAC)  using its fine
 granularity.
% Geometrical coordinates of active streamer tubes in the
% calorimeter are used as inputs for the reconstruction package, which
% builds from geometrical points track candidates. The best track
% hypothesis expressed in terms of the number of hits and corresponding
% $\chi^2$ defines the muon track. Several conditions have been applied on
% the reconstruction to ensure only well defined tracks are used for subsequent
% analysis: the minimal track length is required to be higher
% then $50$ $cm$ and  more then $40\%$ of streamer tubes between first
% and last point of the track are active.

 The
 ability of the accelerator experiment to record cosmic events was
 given by special cosmic trigger defined by the coincidence of $3$
 active sectors in the Time of Flight detector (TOF). The run
 selection was done according to the trigger conditions and all runs
 with the cosmic trigger correctly implemented have been chosen for
 the analysis. The consistency between different trigger
 configurations has been checked in terms of the event rate.
% For
% track multiplicity higher then $3$ all possible trigger configurations give
% consistent event rate around $0.4\ 10^{-2}\ s^{-1}$.

\begin{figure}[h]
%\rule{5cm}{0.2mm}\hfill\rule{5cm}{0.2mm}
%\vskip 2.5cm
%\rule{5cm}{0.2mm}\hfill\rule{5cm}{0.2mm}
\begin{center}
\psfig{figure=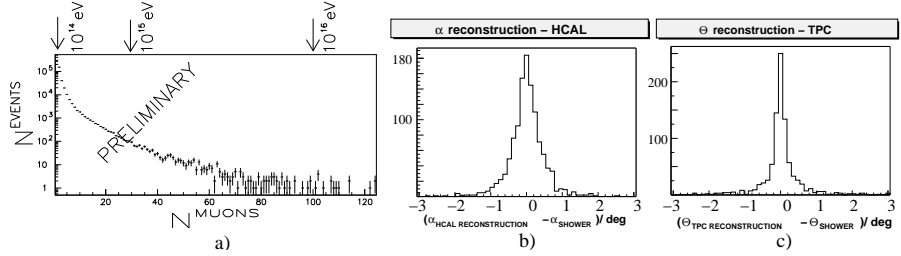,scale=0.14}
%delmu2.eps,scale=0.3}
\end{center}
%height=0.5in}
\caption{a) Muon multiplicity as measured by DELPHI hadron calorimeter
  during $\sim 1.6\ s$ data taking time; b) The difference between
  generated and reconstructed shower angle from HAC in the test MC
  sample; c)~The~difference between
  generated and reconstructed shower angle (zenith) from TPC
\label{fig:multi}}
\end{figure}

 The muon multiplicity distribution as measured in the HAC is plotted
 in Fig.~\ref{fig:multi} a).  
 The values of primary particle energies that are needed to produce showers
 with corresponding multiplicities are shown in the upper part of this
 picture.% They have been obtained using QGSJET model
% for central showers initiated by protons.

 Additionally to events plotted in Fig.~\ref{fig:multi}~a) $7$ events with
 saturated hadron calorimeter have been found in the data sample. The
 number of anode hits from muon chambers has been used  to
 extrapolate multiplicities for $2$ saturated events. There are
 indications that  multiplicity is higher then $150$ in both cases. 

%\section{Simulation}
 The following describes the status of the Monte Carlo (MC) simulation
 in the time of the presentation. Production of large MC data sets is 
 in progress.

 The interaction model ( QGSJET ) is implemented to the
 CORSIKA~\cite{cor}  simulation
 package  which simulates passage of particles through
 the atmosphere and shower development. The rock above the DELPHI
 detector is represented by $5$ layers of materials with different
 densities in a simple GEANT geometry.
 Full simulation of the detector response is used.
  Shower centres are smeared over circular area ($R=200m$) around the
 DELPHI detector.  Large data sets are simulated for proton and iron primary
 particles in the energy range $10^{14}$~-~$10^{18}$~$eV$. Comparison
 between data and MC could show the expected trend to heavier elements
 in the primary particle composition~\cite{kas}.
 The independent check of the relevance of the interaction model will
 be provided.

 The test sample of proton central showers ($\sim 1000$  showers with 
 $E=10^{15}\ eV$) has been used to check the quality of reconstruction
 programs. The differences 
 between generated and reconstructed angles of shower directions
 are plotted in Fig~\ref{fig:multi}~b) and c) for
 HAC and TPC reconstruction.
% As a sample for testing of reconstruction programs about $1000$
% showers with energy $10^15$~$eV$ is used. The primary particles are
% protons and showers are  central with respect to the
% detector, the zenith angle is varied between $0^o\ -\ 60^o$. 
% The difference between  generated
% zenith angle and its reconstructed value from TPC  is plotted in
% Fig... a). The similar comparison for the HAC
% reconstruction in terms of the reconstructed angular projection
% $\alpha$ is shown in Fig. ...b)
%Main principles are described and
% good quality of reconstructed muon tracks is shown at the test sample of
% simulated events.
\section{Conclusions}
 The cosmic multi-muon data are analysed at the DELPHI experiment. The
 distribution of muon multiplicities is measured. Simulation of large
 MC data sets is in progress. MC predictions expresed in
 terms of multiplicity distributions for pure proton and
 iron composition are to be compared with the data in order to search
 for possible excess in the range of high multiplicities. The
 increasing fraction of heavier elements as a function of energy would
 apear in the data with increasing multiplicity as a transition 
 from the proton MC curve to the iron one.
   
{\it  The work was supported by the Ministry of Education of the Czech
 Republic within the projects LN00A006 and LA134.}

\end{document}